# DYNAMIC TRANSPARENT GENERAL PURPOSE PROCESS MIGRATION FOR LINUX

Amirreza Zarrabi

Department of Computer and Communication Systems Engineering, Faculty of Engineering, Universiti Putra Malaysia, 43400 UPM Serdang, Selangor, Malaysia
gs23830@mutiara.upm.edu.my

## ABSTRACT

*Process migration refers to the act of transferring a process in the middle of its execution from one machine to another in a network. In this paper, we proposed a process migration framework for Linux OS. It is a multilayer architecture to confine every functionality independent section of the system in separate layer. This architecture is capable of supporting diverse applications due to generic user space interface and dynamic structure that can be modified according to demands.*

## 1. INTRODUCTION

The primary reason to transfer a program between machines in a network is increasing performance [1]. It introduces an intuitively appealing approach which facilitates dynamic load distribution, fault resilience, easy system administration, and data access locality. It can used to implement various configurations of multicomputer. The literature [2], [3] reveals that in all existing implementations, the process migration mechanism is already hard-coded inside the process migration system. Consequently, all processes would undergo similar phases during the migration event irrespective to the requirement of migration system applications or intrinsic features of the processes.

Furthermore, compatibility and extensibility are considered as fundamental characteristics of a general purpose facility. To date, none of the process migration systems are claimed to be a general purpose implementation as they are extremely dependent on underlying OS and hardware architecture. Support for third party modules or dedicated hardware cannot be simply incorporated inside the migration events and the processes accessing these resources are simply considered as unmigratable.

The optimized migration algorithm deployed in migration event has direct effect on the performance of the process migration system. The literature is inundated with many researches trying to improve the effectiveness of migration algorithms. However, every design has preference factors results in concentration on specific aspect of the migration algorithm [4].

This research is accomplished in two different perspectives, including: architecture and performance. We adopted the following objectives:

- Define a process migration system architecture to segregate the process migration mechanism from the system design so that the minimum flexibility is provided for users to adopt the migration event according to their constraints and to provide the capability of dynamically extending the process migration system on demand to increase compatibility and reduce the system overhead.
- Improve performance of the process migration system by implementing a process migration algorithm, which attempts to integrate the significant features of the existing algorithms to form a generic algorithm, and exploiting the concurrency in process migration system and establishing the notion of parallelism for transferring of the shared states to reduce state relocation time.
- Evaluated the desired characteristics and performed the respective experiments to assess the performance.

## 2. BACKGROUND

A multicomputer configuration is any arrangement of several computers, which is used to support specific services or applications [5]. Any multicomputer configuration could rely on the process concept as a means of distributing work among different computers in the network.

### 2.1. Process Migration

Two major types of process migration exist, including: conventional process migration and migration utilizing distributed resources. Resources on different machines are local in conventional process migration. Consequently, some methods are required to access the remote resources, e.g. using Global Memory Service (GMS) to access the memory pages on remote machines [6]. Conversely, distributed resources belong to the universal reservoir. The process resources are uniquely identified and accessible throughout the network. Therefore, process migration can be easily implemented as the resource distribution would be used to access the migrated process states, e.g. Distributed Shared Memory (DSM) provides the shared memory paradigm so that a migrated process can access to its address space and open files after migration [7].

The granularity of process migration is directly dependent on the extent of remote resource abstraction on the machine, e.g. in DSM shared pages are migrated or replicated in the machine memories, and coherence protocols are deployed to manage the coherency of multiple copies of the same page [8]. However, the GMS coherence semantics for shared pages are the responsibility of the higher-level software that creates sharing in the first place. Therefore, individual threads can be migrated successfully in the migration system based on DSM while the migration system with GMS limits the migration to distinct processes.

### 2.2. Process Migration Algorithms

Whenever transferring a process from the source machine to the destination machine, some steps should be taken, which are similar in all implementations irrespective to the type of migration algorithm and the underlying OS [9]:

- The execution of the process on the source machine is suspended.
- The state of the process is transmitted from the source to the destination machine.
- The state of the process is reconstructed on the destination machine.
- The execution of the process is resumed on the destination machine.
- Information about the process is removed from the source machine.

There are two types of process migration algorithms, including: basic and compound migration algorithms. Compound algorithms are generated by combining the features from the basic algorithms, e.g. assisted post-copy algorithm [10] is a synthesis of the pre-copy [11] and post-copy [12] algorithms. The process algorithms differ in both the process state transfer order and the period that the process is suspended. Consequently, each of these algorithms would have unique characteristics that make them ideal for specific applications. We try to implement the generic migration algorithm proposed in [4] to adhere to the concept of general purpose tool.

## 2.3. Process Migration Mechanism

The method deployed to handle the process states throughout the migration is referred to as the migration mechanism. The process state amount and diversity are increasing as OSs are developing result in complex process migration mechanism as more peculiarities should be considered for a successful migration event. The process state should be treated in one of the following three methods while the process migration is in progress [13]:

- The process state could be ignored on the source machine and simply use the corresponding state on the destination machine.
- The process state could be transferred from the source machine to the destination machine by extracting the required information on the source machine and reinstate them on the destination machine.
- The process state could remain on the source machine and the destination machine would forward the request for operation back to the source machine.

## 3. GENERIC SOFTWARE ARCHITECTURE

Architecture is roughly the prudent partitioning of a whole system into parts, with specific relations among the parts [14]. The architectural structure of a system is important to meet its development, behavioural, and quality goals. Separation of Concerns (SoC) is a de facto standard for architectural designs, which traditionally achieved through modularity of programming and encapsulation based on features or behaviours. Therefore, identifying distinctive aspects of a system as different concerns is a preliminary step in a system design.

Three entities are involved in any migration event: a machine that the process was initially running, a machine that the process resumes execution on, and a state relocation facility. Consequently, three distinct components are introduced to the system:

- Checkpoint/Restart Subsystem.
- Migration Coordinator on source machine.
- Migration Daemon on destination machine.

## 3.1. Checkpoint/Restart Subsystem

The checkpoint/restart subsystem exists on both source and destination machines and accommodates the fundamental property of exporting and importing process states from/to the underlying OS. It is designated to be at the lowest layer in the system architecture with the expectation of accessing to all process resources. It should be capable of masking the intricacies of its duties by offering simplified, general interface to other components.

In any migration event two checkpoint/restart subsystems are participating, one instance checkpoints the process while the other restores the process simultaneously. The contemporary definition of checkpoint and restart, which relies on the concept of separation of the checkpoint and restart events, should be adapted. Each peer should be able to divert the remote machine from its ordinary execution sequence by employing specific protocol.

## 3.2. Migration Coordinator/Daemon

The checkpoint/restart subsystem provides infrastructure for process migration. However, the event could not be initiated unless the source and destination machines agree on certain prerequisites. The migration coordinator derives all desired arguments for the event and invokes the checkpoint/restart subsystem interface to carry out the requested event.

The counterpart to the migration coordinator on the remote machine is the migration daemon. The migration coordinator instructs the daemon for initializing an appropriate environment for migration. The migration daemon establishes equipment for the coordinator to call the checkpoint/restart subsystem interface on the remote machine. They are the trivial components of process migration system and based on system deployment their responsibilities could be assumed and incorporated with other sections of the facilities that employ the checkpoint/restart subsystem.

## 3. SYSTEM DESIGN

When drawing system architecture many design decisions remain unbound and are left to the downstream designers and implementers. Level of implementation is an important aspect of system design, which is not specified in the architecture. Various levels of implementation would result in different performance, complexity, and transparency.

These characteristics contradict with each other and cannot all be satisfied in an individual design. Consider transparency as a principal preference, kernel level implementation is a very apt choice meanwhile system complexity would be adversely affected. The practice of integrating part of the system in user space would offer great flexibility for performing convenient tasks using existing user space tools but with the cost of losing transparency. The optimum system design could be constructed by combining different implementation levels in the process migration system.

## 3.1. Kernel Level Infrastructure

The checkpoint/restart subsystem is selected to be a kernel level infrastructure as it should abide by the behavioural constraints which are only feasible in kernel space. We tried to identify existing similarities in previous checkpoint/restart implementations [15], [16] and proposed a multilayer structure. Each checkpoint/restart system is comprised of at least four layers:

- Kernel subsystem specific layer.
- Transfer medium layer.
- Core system control layer.
- User space interface layer.

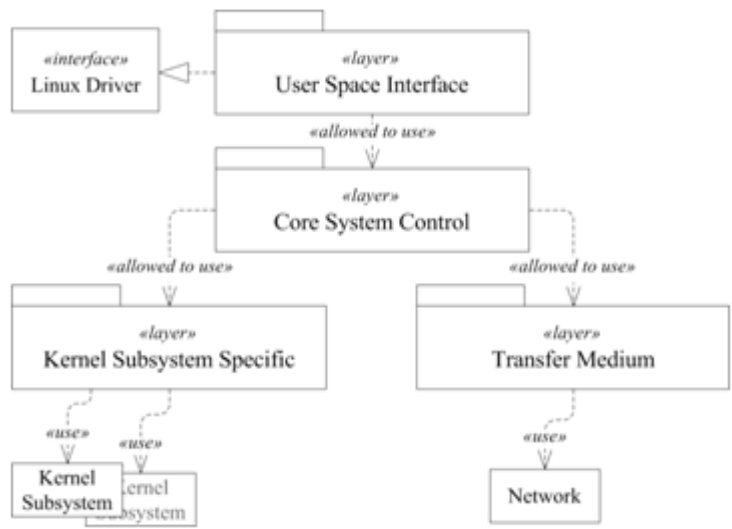

Figure 1. Process Migration System Package Diagram.

A hierarchical structure is defined, which logically partitions the system functionality. Figure 1 presents the organization of system structure in a UML package diagram. Various layers of the system are stacked with the core system control layer as a hub in the middle. Each layer provides service to upper layer either directly or indirectly, e.g. the transfer medium layer can be utilized directly by core system control layer whereas the kernel subsystem specific layer could use the service of the transfer medium layer indirectly through core system control layer.

The layered approach offers several advantages. From developer point of view, system implementation/debugging is easier by separating system functions through a divide and conquer methodology. Communication protocol is generally simpler to design and modify in a layered architecture as the modular nature permits the kernel subsystems or third party modules to store their respective state with native formats.

### 3.1.1. Kernel Subsystem Specific Layer

Linux kernel consists of various subsystems, which are directly or indirectly interacting with the context of processes [17]. Exporting the state of a process implicitly indicates the need for extracting the state of all these subsystems so that while restarting the image of the process can be reconstructed. Type of subsystems recognized by the checkpoint/restart subsystem has an immediate impact on type and number of supported processes. Any kernel subsystem or third party module which obtains an exclusive state should inform the underlying checkpoint/restart subsystem. Consequently, during any event, their state would be considered significant.

The kernel subsystem specific layer resorts to call-backs as a method to obtain enough knowledge on management policy of each subsystem. As a result, each kernel subsystem or third party module has to register predefined set of operations in this layer. These operations would direct the kernel subsystem specific layer to store and retrieve subsystem specific state. Refer to Table 1 for the proposed list of operations.

Table 1.  Kernel Subsystem Operation.

| Operation | Description |
|---|---|
| *checkpoint* | Checkpoint the state of kernel subsystem or third party module. |
| *restart* | Restart the state of kernel subsystem or third party module. |
| *fault* | Recover an entity[a] from error state. |
| *status* | Extract the current status of ongoing event. |
| *dump* | Handle fatal error incidents and printing error information. |

a. Entity refers to an element used as a unit in checkpoint and restart event, e.g. a memory page is considered as an entity for memory subsystem.

The *fault* method would be invoked to divert the checkpoint or restart event from the normal execution sequence if required, and it is sensible, provided that they could be conducted in multiple steps. Therefore, the *checkpoint* and *restart* methods should be iterative type. That is, a single call to these methods could decline to fulfil the respective event.

Moreover, the shared resources, managed by the kernel subsystems or third party modules, should undergo the checkpoint or restart event with extra considerations. They require to be checkpointed once for all the processes that reference the resources and should remain in the consistent state while a checkpoint event is in progress. Processes should be restarted while referencing to their formerly shared resources [16].

### 3.1.2. Transfer Medium Layer

The transfer medium layer specifies how the extracted states from kernel subsystems or third party modules should be handled. The primary objective of this layer is to decouple low speed transfer medium from the checkpoint and restart events. However, different types of transfer medium could be deployed, ranging from a standard filesystem interface of particular OS [18] to a specifically optimized network protocols [19]. Irrespective to the characteristics of transfer mediums, they should provide a facility to store and retrieve data. In transfer medium layer, call-backs are employed to abstract the peculiarities of a transfer medium. The minimal operation that a transfer medium module should define is represented in Table 2.

Table 2.  Transfer Medium Operations.

| Operation | Description |
|---|---|
| *popdata* | Retrieve medium data into kernel subsystem or third party module. |
| *pushdata* | Store kernel subsystem or third party module data in the medium. |
| *command* | Send or receive medium dependent command. |
| *flush* | Flush all buffers in the transfer medium. |

Conventionally, the transfer medium would operate with a portion of optimizations, which could end in deadlock in particular occasions. Optional operations should exist to asynchronously manipulate optimizations. The *command* method is used to send low-level instructions to transfer medium layer in order to modify its behaviour, e.g. changing the internal buffer size. The same method can be exploited by kernel subsystems or third party modules to send their internal command to peer modules on the remote machine if the network medium is used.

### 3.1.3. Core System Control Layer

Core system control layer is glue that keeps the whole system together. All process specific information concerned with handling a Linux multithreaded process within the checkpoint/restart subsystem are maintained at this layer, e.g. process specific allocated buffers or remote machine peer process identifications. The checkpoint and restart events on multiple processes are kept thoroughly isolated in this layer and do not interfere with each other. Therefore, the concurrency of events is guaranteed while the kernel subsystem specific layer takes care of shared resources.

The core system control layer should not make any assumption about specific process undergoes an event. As a workaround, each process is treated as an object with its own constructor/de-constructor functions which are specified based on request attributes. Table 3 shows the operations utilized for each process [20].

Table 3. Process Specific Constructor/De-constructor.

| Operation | Description |
|---|---|
| *setup* | Called before launching checkpoint/restart to prepare the process |
| *restart* | Called when restarting a process |
| *cleanup* | Called for terminating current success/failed event on a process |

The actual checkpoint or restart is performed within this layer. As a result, an execution context should be initiated to carry out the checkpoint or restart event. Contrary to Linux specification which confines each process to access and modify its own resources, the execution context should assure consistent access to arbitrary process.

### 3.1.4. User Space Interface Layer

This layer introduces a single entry point to the checkpoint/restart subsystem that accepts a request for a particular event from a user, evaluates the request and issues the required sequence of calls to core system control layer. In other words, it is a component of the system that translates a user action into one or more requests for system functionality and feedback about the consequences of this action [21].

### 3.1. User Level Tools

The user level tools refer to user space applications that deploy the checkpoint/restart subsystem. They are trivial components and most of their operations could be combined with any user applications that rely on checkpoint/restart subsystem. User level tools, irrespective to whether they are implemented as standalone migration coordinator/daemon or embedded inside normal user applications, should establish an appropriate environment on the remote machine and issue series of requests to underlying checkpoint/restart subsystems.

The checkpoint/restart subsystem attempts to isolate the processes concurrent events. Many applications consist of multiple cooperating processes so that not only must the state associated with each process be migrated, but the processes relationships must be considered. This compels the requirement of providing an extra level of synchronization in user level tools when an event is in progress for related processes.

## 4. SYSTEM IMPLEMENTATION

Process migration system scalability was one of the concerns while designing the system. To meet this objective, the kernel subsystem specific layer and the transfer medium layer were designed with a manager as a coordinator that supplies certain services to insert and remove kernel subsystems and transfer medium contexts to the checkpoint/restart subsystem. This structure is depicted in Figure 2.

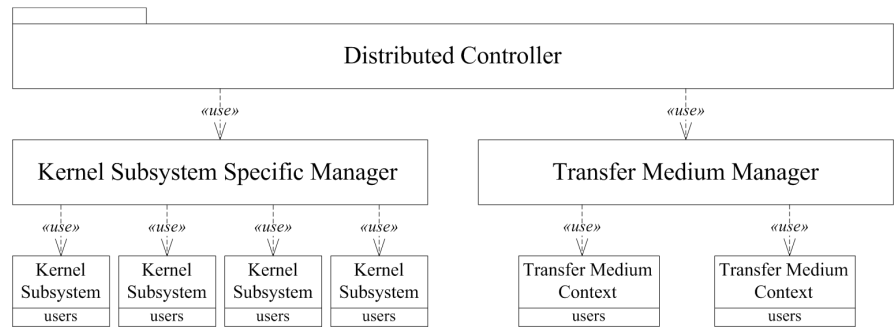

Figure 2. Distribution Architecture of Migration System.

Fundamental to any distributed systems is the concurrency and collaboration among multiple machines. In many cases, this also means that same resources should be manipulated simultaneously. A distributed synchronization level is required to preserve system consistency and facilitate inserting/removing kernel subsystems and transfer medium contexts on-fly. The distributed controller would be implemented as an independent user space tool on all machines in the network above the process migration system and orchestrates these events.

The execution context for checkpoint and restart events is initiated inside the core system control layer. The execution context could be external or internal according to the user request submitted to the checkpoint/restart subsystem. In the internal context, the respective event would happen within the context of the process itself, contrary to the external context which the event would be accomplished within a different process context [22].

Irrespective to the type of execution context, it has to assure consistent access to the target process resources using standard Linux functions. While the internal execution context would have straight access to the process resources by default, the external execution context needs extra consideration. The worker method related to the requested event would be invoked in the execution context.

### 4.1. Checkpoint and Restart Worker Methods

Particular worker methods are devoted to checkpoint and restart events. These methods rely on kernel subsystems or third party modules and transfer medium contexts to carry out the corresponding events. They would satisfy the requirements of the event by calling the process constructor stored in setup method. The restart worker method tries to determine if a process is willing to restart and would call the restart method ahead of resuming execution of the process. This is only valid if restart happens using an external execution context. Likewise, the events are concluded by calling the cleanup method.

## 4.2. Generic Internal Watchdog

Watchdog or Computer Operating Properly (COP) timer is a software timer that triggers a corrective action if the main program neglects to operate regularly due to some fault condition. The intention is to bring the system back from the unresponsive state into normal operation or generate appropriate error. A separate watchdog timer is instantiated for each process which its respective call-back would oversee the progress of the event.

The event is considered frozen if the checkpoint and restart methods of all contributing kernel subsystems or third party modules do not declare a success state for the specified period of time. The delineation of the success notion is kernel subsystem or third party module specific and depends on whether the last invocation assisted progress of the event even if an error incident was handled in previous call. The internal watchdog plays a significant role in the system consistency and deadlock detection, e.g. error originated from transfer medium context could not be published to the remote machine while the transfer medium is unreliable and the internal watchdog should handle the erroneous condition.

## 4.2. Process Migration Synchronization

The checkpoint and restart methods of kernel subsystems or third party modules are iterative type. As a result, the checkpoint and restart worker methods may invoke them several times until they report the operation completion. The execution sequence should be synchronized between source and remote machine while the event is in progress.

Ordinarily, the checkpointing machine is the data sender while the restarting machine in the data receiver. The communication channel established between the source and destination machine operates in the simplex mode. However, the destination machine should inform the source machine about the previous step so that it could proceed posting new data. The command method of transfer medium module is exploited to send the intended message to the specific kernel subsystem or third party module in the source machine.

## 4.2. User Space Interface

The interface between the kernel space and the user space is implemented as character device driver. Character devices are accessed through names in filesystem. Those names are called special files or device files and are conventionally located in the /dev directory [23]. The ioctl system call is exploited for interaction with the device file. It multiplexes the different commands to the appropriate functions in the user space interface layer.

Ordinarily, in any repetitive user data access, the system should have some means of establishing context [24]. In the user space interface layer, each request is an abstract that constructs the context of the event. Any request is identified with an integer value, which is returned to the user space when a request is submitted. While the request ID represents a submitted request locally, a dedicated token is used for distributed representation of an event. It maps the request ID from the source machine to the destination machine. For each event, an instance of this token should be created, which is used in the handshaking sequence. The handshaking is an automated process of negotiation that dynamically sets parameters of a communications channel established between source and destination machines before normal communication over the channel begins.

## 4.2. Multithreaded Process State Consistency

The kernel subsystems and third party modules require consistent access to all process states. However, a running process would modify its states as a part of the execution side effect, and some of these states are only available in particular process execution modes, e.g. processor register contents are available in the kernel during exceptions, system calls, or interrupts.

Therefore, the core system control layer should temporarily render the process inactive at specific points in its execution to achieve the requested consistency. The user space interface layer provides two methods to quiesce a process, including: synchronous and asynchronous.

In synchronous method, all threads inside a process should collaborate to quiesce the process. All threads of the process should call a specified system call to force entering to the kernel space where the thread would be blocked. However, the asynchronous method would exploit Linux freezer subsystem.

## 4.2. Process Migration Event Strategies

A process migration event strategy refers to one of the options that a user can choose to checkpoint and restart a process on different machines. Depending on the execution context type in the core system control layer and quiescing method, various strategies could be proposed for migration of a process in the user space interface layer. Not every combination of the execution context and quiescing method is valid for process migration because of the checkpoint/restart subsystem weakness in mapping process threads on the source machine to the respective threads on the destination machine and the kernel restriction on direct access to context of processes from outside.

## 5. EXPERIMENTAL PLATFORM

The platform consists of two physical and two virtual machines. The architecture of the test bed is depicted in Figure 3. Underlying network is comprised of two segments:

- Actual segment which is used to connect two physical machines
- An extension segment which is simulated using Virtual Network Editor bundled into VMWare Workstation.

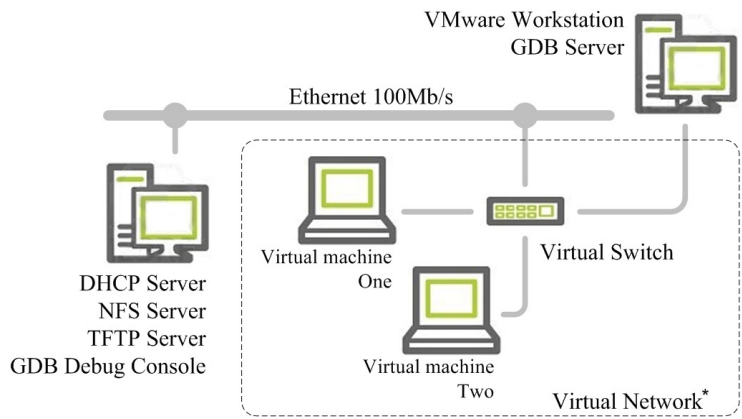

Figure 3. Test Bed Architecture.

These two segments are interconnected using a virtual switch. The two virtual machines are defined to be diskless which are booted over the network using Remote Network Boot via Preboot eXecution Environment (PXE). The root filesystem is published using NFS over the network which is mounted by the kernels while booting. The main server assumed the role of Dynamic Host Configuration Protocol (DHCP), Trivial File Transfer Protocol (TFTP), and NFS server to offer these functionalities. The measurements were conducted on two Fedora 9 OSs with kernel 2.6.30 running on virtual machines.

## 5. MIGRATION SYSTEM

As the address space of a process constitutes the largest part of process state, the process migration latency time is calculated exploiting a single process with various address space sizes. Figure 3 shows the latency time with different address space sizes in the proposed migration system.

The latency time is proportional to the size of the process address space and would increase almost linearly with its growth. The characteristic of the process has a direct effect on migration latency time. The processes with a higher memory footprint would bear more latency time as more dirty pages would be produced either in memory or swap space, which should be transferred. As depicted in Figure 3, the migration latency time for a process without memory footprint would decline by almost 80% with 9MB of address space compared to the similar process with memory footprint.

The proposed migration system support for the early event request could compensate the long migration latency time. The process address space transfer can be broken down into two phases: process page table scan and dirty page checkpoint and restart. The total time spent by two phases yield major part of the process migration latency time. The early event request would eliminate these overheads by enabling laziness. The outcome from this approach is represented in Figure 3. The latency time for processes with memory footprint is decreased by average of 0.47 Sec. while it is limited to 0.05 Sec. for processes without memory footprint.

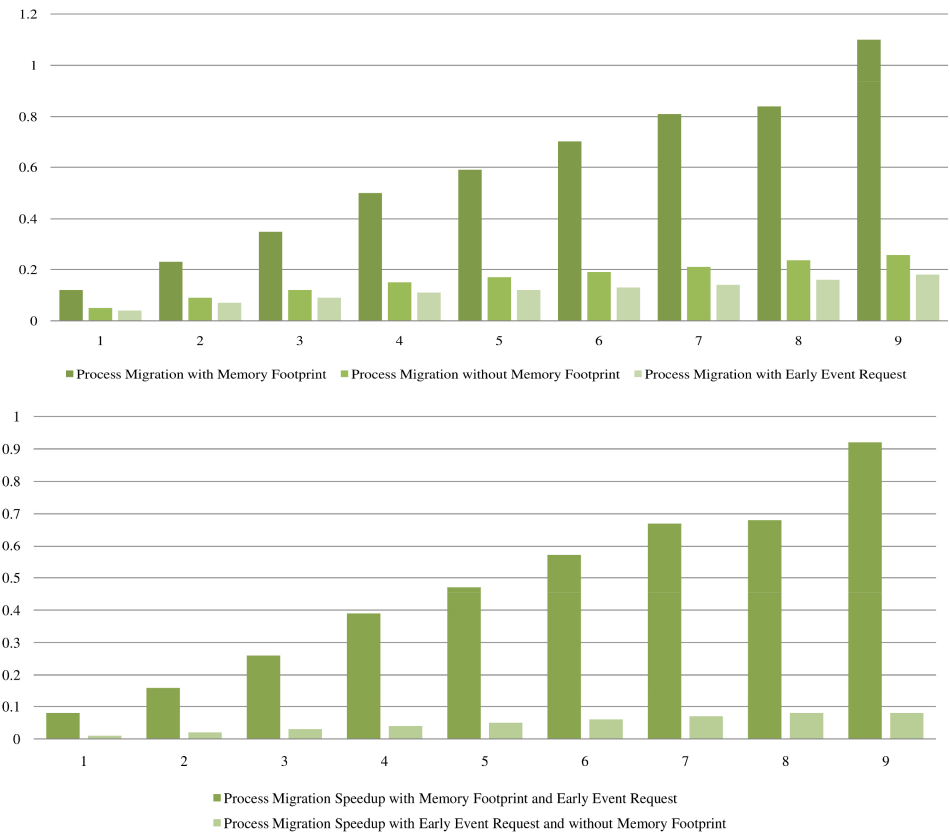

Figure 3. Process Migration Latency Time and Different Address Space Size

The migration speedup for processes with memory footprint is increased from 65% for a 1MB of address space to 80% for 9MB of address spaces. This is due to insignificant overhead of page table traversal as well as checkpoint and restart of pages in a small memory region. This instructs the fact that, if the proposed migration system is utilized for any load balancing solution to free up memory resource, then a long-running process with large address space could be an appropriate candidate for the migration event.

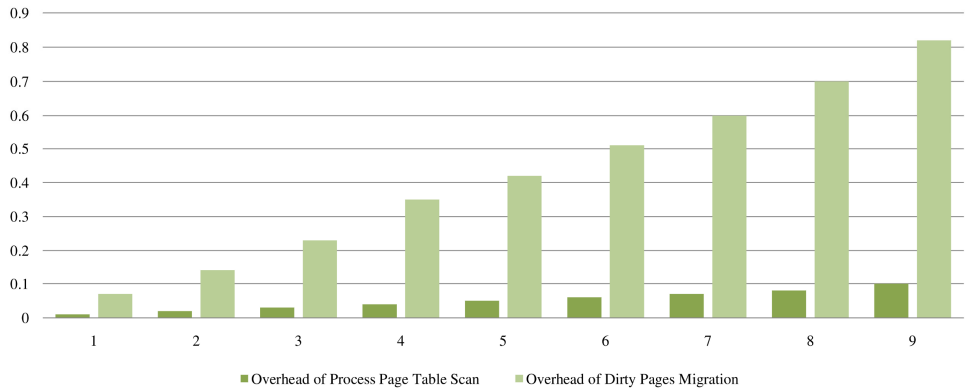

Figure 4.  Overhead of the Address Space Migration.

The migration speedup for processes without memory footprint is smaller in a great amount compared to the processes with memory footprint. The processes without memory footprint would not participate in checkpoint and restart of memory pages as a result using the early event request solely affects the process page table scanning overhead. The overhead of process address space transfer is depicted in Figure 4. The overhead of dirty page migration for a process with 9MB of address space is almost eight times more than the overhead of the process page table scan.

Figure 5 demonstrates migration elapsed time for multiple processes with 1MB of address space. The coordinator negotiation overhead is bounded to time consumed by the handshaking sequence in the first phase of the process migration. This time raised almost linearly with the number of processes undergo the migration event as reconstructing a process frame in remote machine constrains collecting data for each process. This routine would be repeated for every migrating process results in an upturn in negotiation overhead.

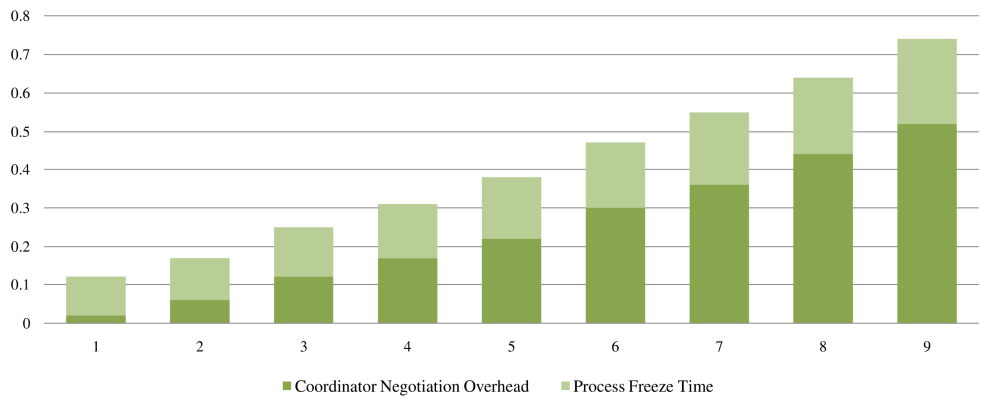

Figure 5.  Process Migration Event Time Consumption

The process freeze time is almost constant independent to the number of processes as events are concurrent. Some factors may perhaps insert the slight increase in freeze time contrary to expected behaviour. The coordinator would initiate the migration events one after the other for every process whereas the simultaneous events are proceeding concurrently. Quiescing a process exploiting the Linux kernel freezer subsystem occurs when execution control transferred from kernel space to user space. Therefore, the current machine CPU load and scheduling policy would disturb freeze time. Similarly, the network traffic could affect the freeze time of processes.

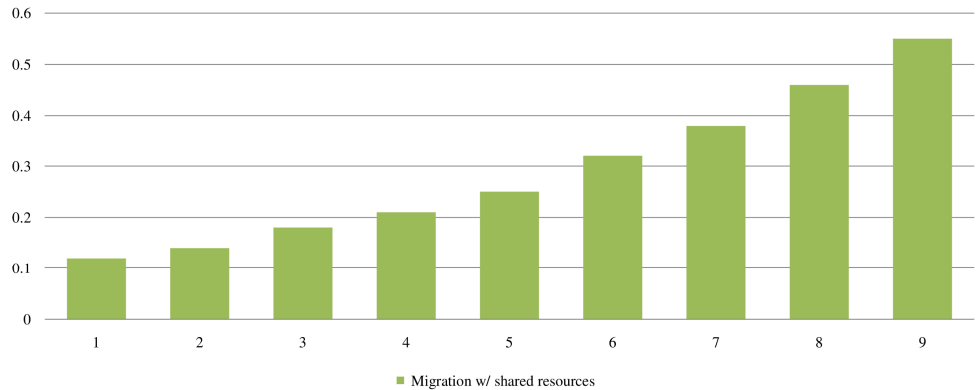

Figure 6. Process Migration Event Time Consumption

The proposed migration system tries to exploit the process migration concurrency which is utilized for reducing the freeze time as a means of cutting down the state relocation time. The shared resources would be identified and the role of relocating shared resources is distributed between all participating processes resulting in data consistency and migration speedup. This is an evident effect as all shared resources would be only checkpointed once and with collaboration of all processes undergo the event concurrently.

Figure 6 illustrates the migration of multiple processes with shared memory regions in their address space. Comparing results with data obtained from Figure 5 demonstrates that existence of shared resources and the parallelism deployed by the system could end up process migration speedup. It is tangible that for the migration of nine processes, the speedup of around 20% is achieved. The system cannot obtain any speedup for migration of one process as we expect.

## 3. CONCLUSIONS

In this paper we presented a process migration system for Linux OS. Concentration is devoted to the architecture, design and implementation. No assumption is made for the underlying OS or hardware architecture. The characteristic of the whole system is defined by the subset of kernel subsystem or transfer medium modules loaded in to the framework. It allows users to uniquely identify the kernel subsystems or third party modules so that the process migration mechanism can be specified by the user while a request for an event is submitted to the checkpoint/restart subsystem.

This architecture imposes no limitation on the implementation of the kernel subsystem of transfer medium module as it supports the concept of migration event strategies. Therefore, the same framework can be exploited for implementation of the checkpoint/restart system.

# REFERENCES

[1] A. S. Tanenbaum and M. Steen, *Distributed systems: principles and paradigms*: Pearson Prentice Hall, 2007.

[2] D. S. Milojicic, *et al.*, "Process migration," *ACM Computing Surveys,* vol. 32, pp. 241-299, 2000.

[3] R. Lawrence, "A survey of process migration mechanisms," *Dept. of CS, Univ. of Manitoba, May,* vol. 29, 1998.

[4] A. Zarrabi, "A Generic Process Migration Algorithm," *International Journal of Distributed and Parallel systems,* vol. 3, September 2012.

[5] S. D. Burd, *Systems Architecture*: COURSE TECHNOLOGY, 2010.

[6] M. J. Feeley, *et al.*, "Implementing global memory management in a workstation cluster," in *ACM SIGOPS Operating Systems Review* vol. 29, ed, 1995, pp. 201-212.

[7] G. Vallee, *et al.*, "Process migration based on gobelins distributed shared memory," in *Cluster Computing and the Grid, 2002. 2nd IEEE/ACM International Symposium on*, ed, 2002, pp. 325-325.

[8] C. Morin, *et al.*, "Towards an efficient single system image cluster operating system," *Future Generation Computer Systems,* vol. 20, pp. 505-521, 2004.

[9] M. Richmond and M. Hitchens, "A new process migration algorithm," *ACM SIGOPS Operating Systems Review,* vol. 31, pp. 31-42, 1997.

[10] M. Noack, "Comparative evaluation of process migration algorithms," *Master's thesis, Dresden University of Technology-Operating Systems Group,* 2003.

[11] M. M. Theimer, *et al.*, *Preemptable remote execution facilities for the V-system* vol. 19: ACM, 1985.

[12] R. S. C. Ho, *et al.*, "Lightweight process migration and memory prefetching in OpenMosix," in *Parallel and Distributed Processing, 2008. IPDPS 2008. IEEE International Symposium on*, ed, 2008, pp. 1-12.

[13] F. Douglis, "Transparent process migration in the Sprite operating system," Citeseer, 1990.

[14] P. Clements, *et al.*, *Documenting software architectures: views and beyond*: Addison-Wesley Professional, 2010.

[15] O. Laadan and S. E. Hallyn, "Linux-CR: Transparent Application Checkpoint-Restart in Linux," in *Proceedings of the 12th Annual Ottawa Linux Symposium (OLS), Ottawa, Canada*, ed, 2010.

[16] J. Duell, *et al.*, *The design and implementation of Berkeley Lab's linuxcheckpoint/restart*: Citeseer, 2005.

[17] W. Mauerer, *Professional Linux kernel architecture*: Wrox, 2008.

[18] H. Zhong and J. Nieh, "CRAK: Linux checkpoint/restart as a kernel module," *Department of Computer Science, Columbia University, Tech. Rep. CUCS-014-01,* 2001.

[19] A. Barak and O. La'adan, "The MOSIX multicomputer operating system for high performance cluster computing," *Future Generation Computer Systems,* vol. 13, pp. 361-372, 1998.

[20] W. R. Dieter and J. E. Lumpp Jr, "User-level checkpointing for LinuxThreads programs," in *Proceedings of the FREENIX Track: 2001 USENIX Annual Technical Conference*, ed, 2001, pp. 81-92.

[21] B. A. Myers and M. B. Rosson, "Survey on user interface programming," in *Proceedings of the SIGCHI conference on Human factors in computing systems*, ed, 1992, pp. 195-202.

[22] R. Gioiosa, *et al.*, "Transparent, incremental checkpointing at kernel level: a foundation for fault tolerance for parallel computers," in *Proceedings of the 2005 ACM/IEEE conference on Supercomputing*, ed, 2005, p. 9.

[23] J. Corbet, *et al.*, *Linux device drivers*: O'Reilly, 2005.

[24] S. L. Smith and J. N. Mosier, *Guidelines for designing user interface software*: Citeseer, 1986.

## Author

Amirreza Zarrabi has received his master's degree in Computer Communication and Networks from University Putra Malaysia in 2012. His research interests are in computer architecture, device modelling, embedded systems, operating systems, system security, web services, and distributed computing. During his thesis, he was working on distributed memory, file management, and process migration in Linux operating system. Currently he is working on robust transport protocol for dynamic high-speed networks as a research engineer in an international investment solution provider company in Kuala Lumpur.

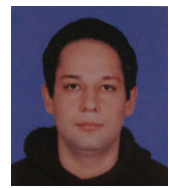